
\magnification=1200
%
%
\def\drho{{\partial _\rho}}
\def\hdrho{{\hat \partial_\rho}}
\def\drhobar{{\partial _{\bar \rho}}}
\def\dz{{\partial _z}}
\def\dzbar{{\partial _{\bar z}}}
\def\pp{{\psi^+}}
\def\pd{{\psi^-}}
\def\ep{{\eta^+}}
\def\em{{\eta^-}}
\def\Dp{{D_{\pp}}}
\def\Dm{{D_{\pd}}}
\def\hDp{{\hat D_\pp}}
\def\hDm{{\hat D_\pd}}
\def\Dep{{D_{\ep}}}
\def\Dem{{D_{\em}}}
\def\pim{{\partial_m x^{\mu}-i\partial_m \theta^{\alpha}\gamma^{\mu}
_{\alpha\beta}\theta^{\beta}}}

\def\tgt{{\theta^{\alpha}\gamma^{\mu}_{\alpha\beta}\theta^{\beta}}}
\def\TgT{{\Theta^{\alpha}\gamma^{\mu}_{\alpha\beta}\Theta^{\beta}}}
\def\dt {{\partial_\tau}}

\def\gmu {{\gamma^\mu_{\alpha\beta}}}
\tolerance=5000
\footline={\ifnum\pageno>1
       \hfil {\rm \folio} \hfil
    \else \hfil \fi}

\overfullrule=0pt 
\baselineskip=18pt
\parskip=12pt 
\raggedbottom
\centerline{\bf The Heterotic Green-Schwarz Superstring}
\centerline{\bf On An N=(2,0) Super-Worldsheet}
\vskip 24pt
\centerline{Nathan Berkovits}
\vskip 12pt
\centerline{Institute for Theoretical Physics}
\centerline{State University of New York at Stony Brook}
\centerline{Stony Brook, NY 11794}
\vskip 12pt
\centerline{e-mail: nathan@dirac.physics.sunysb.edu}
\vskip 12pt
\centerline {December 1991}
\vskip 12pt
\centerline {ITP-SB-91-69}
\vskip 36pt
By defining the heterotic Green-Schwarz superstring action on an
N=(2,0) super-worldsheet, rather than on an ordinary worldsheet,
many problems with the interacting Green-Schwarz superstring
formalism can be solved. In the light-cone approach, superconformally
transforming the light-cone super-worldsheet onto an N=(2,0)
super-Riemann surface allows the elimination of the non-trivial
interaction-point operators that complicate the evaluation of
scattering amplitudes. In the Polyakov approach, the
ten-dimensional
heterotic Green-Schwarz covariant action defined on an N=(2,0)
super-worldsheet can be gauge-fixed to a free-field action with
non-anomalous N=(2,0) superconformal invariance,
and
integrating the exponential of the covariant action over all
punctured N=(2,0) super-Riemann surfaces produces scattering
amplitudes that closely resemble
amplitudes obtained using the unitary
light-cone approach.
\noindent
\vfil\eject
\centerline{\bf I. Introduction}
The Green-Schwarz formalism for the superstring has the
advantage over the Neveu-Schwarz-Ramond formalism of being
manifestly spacetime supersymmetric. Besides eliminating the
need to sum over spin structures, the spacetime supersymmetry
of the Green-Schwarz formalism considerably simplifies the
evaluation of superstring scattering amplitudes involving
external fermions.$^1$

Unfortunately, at the present time, there are only two methods for
calculating Green-Schwarz superstring scattering amplitudes, neither
of which has been very productive. The first method starts from the
light-cone gauge Green-Schwarz action and uses Feynman's
manifestly unitary prescription
of integrating the exponential of the action over all possible
light-cone worldsheets to evaluate the scattering amplitudes.$^{2,3,4}$ Besides
the usual free part of the light-cone gauge action, there is also
an interaction term that is completely determined by Lorentz invariance.
This interaction term is not just an overlap delta function in the
string fields (as in the Veneziano string$^5$), but contains additional
field dependence. After conformally transforming the light-cone
worldsheet to a smooth Riemann surface, these interaction terms become
operator insertions on the surface
at the interaction points of the light-cone
worldsheet. Because it is extremely difficult to determine
the location of these interaction points on the Riemann surface (the
locations depend on the $P^+$'s of the external strings), no
amplitudes involving more than one loop or more than
four external strings have been explicitly evaluated using this method.

The second method for calculating amplitudes starts from the
ten-dimensional covariant Green-Schwarz action
defined on an ordinary worldsheet$^6$ and
uses Polyakov's ansatz$^7$ that integrating
the exponential of this covariant
action over all punctured Riemann surfaces will give the correct
superstring scattering amplitudes.$^{8,9}$ In order to write the covariant
action in terms of free fields, it is necessary to gauge-fix
both the two-dimensional reparameterization invariance
and the fermionic $\kappa$-symmetries. In this semi-light-cone
gauge, the conformal anomaly from the matter fields and ghosts is
non-zero,$^{10}$ implying that the usual method for regularizing the
free-field action
fails to preserve all of the classical symmetries.$^{11}$
In
hindsight, this complication is not surprising since
a free-field action on ordinary Riemann surfaces would be unable
to reproduce the interaction-point operator insertions that
are present in the manifestly unitary light-cone method.

It is worthwhile to analyze how in the Neveu-Schwarz-Ramond formalism
for the superstring, these two methods were used successfully to
calculate scattering amplitudes. As in the Green-Schwarz formalism, the
Neveu-Schwarz-Ramond light-cone gauge action also has a non-trivial
interaction term that leads to operator insertions at the light-cone
interaction points.$^{12}$ However, it was shown that by writing the
Neveu-Schwarz-Ramond light-cone action on an N=1 super-worldsheet, the
interaction term simplifies to an overlap delta function in the
string superfields, thereby eliminating the operator insertions at
the interaction points.$^{13}$ In other words, by considering light-cone
super-worldsheets that could be superconformally transformed to an
N=1 super-Riemann surface (and integrating over the N=1 super-moduli),
one could eliminate any operators that explicitly depended on the
locations of the light-cone interaction points. After this
simplification, the light-cone methods were successfully used to
calculate Neveu-Schwarz-Ramond scattering amplitudes with an arbitrary
number of loops and an arbitrary number of external Neveu-Schwarz
bosons.$^{14,15}$

In the Neveu-Schwarz-Ramond formalism, Polyakov's ansatz
can also be used to evaluate superstring scattering
amplitudes. However, instead of integrating the exponential of the
covariant Neveu-Schwarz-Ramond action over all punctured ordinary
Riemann surfaces, one integrates over all punctured N=1
super-Riemann surfaces.$^{16}$
Since the conformal anomaly vanishes after gauge-fixing the
non-Liouville parts of the super-vierbein, the usual method for
regularizing the free-field action can be used.
This
does not contradict
the light-cone method since after superconformally
transforming the light-cone super-worldsheet to an
N=1 super-Riemann surface, there are no operator insertions at the
interaction points.

In this paper, it will be shown that by replacing the ordinary
two-dimensional worldsheet of the heterotic Green-Schwarz
superstring action$^{17}$ with an N=(2,0) super-worldsheet, the problems
with both the light-cone and covariant methods in the Green-Schwarz
formalism can be solved as they were in the
Neveu-Schwarz-Ramond formalism. That is, after
superconformally transforming the Green-Schwarz
light-cone super-worldsheet onto an N=(2,0) super-Riemann
surface and integrating over the super-moduli, there are no
operator insertions at the light-cone interaction points. Furthermore,
after gauge-fixing all of the symmetries in the ten-dimensional
covariant heterotic Green-Schwarz action except for the N=(2,0)
superconformal invariance, the conformal anomaly from the
remaining matter fields and ghosts is zero.

The first half of this paper will discuss the light-cone method for
evaluating heterotic Green-Schwarz scattering amplitudes. After
reviewing the light-cone gauge heterotic Green-Schwarz action
on an ordinary worldsheet, it will be shown how to define the light-cone
action on an N=(2,0) super-worldsheet. It will then be proven that
integrating the exponential of the light-cone action over ordinary
worldsheets and including the appropriate operator insertions at the
light-cone interaction points is equivalent to integrating the exponential
of the light-cone action over N=(2,0) super-worldsheets without any
operator insertions at the interaction points. Because there is no
longer any explicit dependence on the locations of the light-cone
interaction points, evaluation of Green-Schwarz scattering
amplitudes using this new version of the light-cone method should
be considerably simpler than using the old method.

The second half of the paper will discuss the Polyakov
method for evaluating heterotic Green-Schwarz scattering amplitudes.
It will first be shown that
the ten-dimensional covariant Green-Schwarz action on an ordinary
Riemann surface contains a conformal anomaly in the semi-light-cone
gauge.$^{10}$
It will next be shown how to define the covariant heterotic Green-Schwarz
superstring action on an N=(2,0) super-Riemann surface (this N=(2,0)
action for the heterotic Green-Schwarz superstring was first defined
by Tonin$^{18}$, who relied heavily on earlier work done by Sorokin, Tkach,
Volkov, and Zheltukhin on the superparticle$^{19}$).
After fixing the action in a gauge where all fields are free, the
conformal anomaly is calculated and shown to vanish.
This means that BRST quantization of the interacting
Green-Schwarz superstring should be possible in this gauge,
although it would not be manifestly Lorentz-covariant (the
gauge choice breaks the manifest spacetime SO(9,1) invariance down
to U(4), but preserves the worldsheet N=(2,0) superconformal
invariance).
Finally,
it is argued that the ansatz of integrating the exponential of
the Lorentz-covariant heterotic Green-Schwarz action over all
punctured N=(2,0) super-Riemann surfaces produces scattering
amplitudes that are in agreement (up to an as yet undetermined
measure factor) with amplitudes obtained using the unitary light-cone
method.

At the present time, only the heterotic version of the Green-Schwarz
covariant action is able to be defined on N=2 super-Riemann surfaces.
Although the light-cone method easily generalizes to N=(2,2)
super-worldsheets, and can therefore be used to describe non-heterotic
versions of the Green-Schwarz superstring, it is not yet clear
if the Lorentz-covariant method can be similarly generalized.
\vskip 24pt
\centerline {\bf II.A. The Light-Cone Method on an Ordinary Worldsheet}
The light-cone method for the interacting Green-Schwarz superstring
was first introduced by Green and Schwarz in 1982 to describe
external strings with momenta $P^+=0$,$^1$ and was later generalized
by Green, Schwarz,$^2$ and Mandelstam$^3$ to describe external strings with
arbitrary $P^+$.

The light-cone heterotic Green-Schwarz fields
consist of a bosonic
real SO(8) vector, $x^j(\tau ,\sigma )$, a fermionic real SO(8) chiral spinor,
$s^a(\tau ,\sigma )$, and 32 fermionic
real SO(8) scalars, $\phi ^p(\tau ,\sigma )$,
that parameterize the self-dual lattice of the heterotic string.$^{17}$

The free light-cone gauge action for these fields is
$$\int d\tau d\sigma [ \drho x^j \drhobar
x^j +i s^a \drhobar s^a +i \phi ^p \drho
\phi ^p ]\eqno(II.A.1)$$
where
$\rho \equiv \tau +\sigma$ and $\bar \rho \equiv \tau -\sigma$.
Although this action uses a two-dimensional Minkowski metric, it
is straghtforward to Wick-rotate to
two-dimensional Euclidean space, in which case
$\rho \equiv \tau +i\sigma$, $\bar \rho \equiv \tau -i\sigma$, and the
fermionic fields, $s^a$ and $\phi ^p$, change from being real-valued
to being complex-valued (in the heterotic superstring, the complex
conjugates of $s^a$ and $\phi ^p$ are treated as linearly
independent variables, and are set to zero).

In order to preserve the global spacetime supersymmetry of the
action under $s^a (\tau , \sigma ) \rightarrow s^a (\tau , \sigma ) +
\epsilon ^a$, $s^a$ must be a periodic function everywhere on the
light-cone string worldsheet (note that
$\phi ^p$ can be either periodic or anti-periodic).
Therefore, since $s^a (\tau,\sigma)$ transforms like a worldsheet
spinor under conformal transformations, $\hat s^a ( z, \bar z )
= \sqrt {\dz\rho }\, s^a (\rho ,\bar \rho ) $
has square-root cuts at the zeroes and poles of $\dz\rho
$, where $\rho (z)$ is the conformal
transformation that maps the light-cone interacting string worldsheet
onto a smooth punctured Riemann surface with complex coordinates,
$z$ and $\bar z$.
These zeroes and poles of
$\dz\rho $
occur at the interaction points of the light-cone worldsheet and at
the ends of the external strings. Since these square-root cuts can
occur in either the numerator or denominator of $\hat s^a$, it is
necessary to split the eight $s^a$'s into four complex pairs,
$s^{+l} \equiv s^l +i s^{l+4}$ and $s^{-\bar l}\equiv  s^l -i s^{l+4}$
for $l$=1 to 4,
and to choose boundary conditions for the ($s^{+l}, s^
{-\bar l}$) fields at the interaction points and at the ends of the
external strings. For example, choosing the boundary condition that
$\hat s^{+l}(z,\bar z)$ goes like $\sqrt {z - \tilde z}$
for $l$=1 to 4 near the interaction point $\tilde z$,
means that the correlation function of $\hat s^{+l} (v,\bar v)$ with
$\hat s^{-\bar l} (w, \bar w )$ will go like
$\sqrt {v- \tilde z}$ when $v \rightarrow \tilde z$, but
like $(\sqrt {w- \tilde z})^{-1}$ when $w \rightarrow \tilde z$.

Once boundary conditions have been chosen for the ($s^{+l}, s^
{-\bar l}$) fields, Feynman's manifestly
unitary prescription of integrating the
exponential of the interacting light-cone action over all
possible worldsheets connecting the initial and final string
states can be used to evaluate S-matrix scattering amplitudes.
As mentioned in the introduction, the interaction term of
the light-cone Green-Schwarz Lagrangian is not just an
overlap delta function in the string fields, but instead
has the following form:$^{2,3}$
$$L_{GS}^{int}=\lambda \delta (x^j_{in}-x^j_{out})
\delta (s^a_{in}-s^a_{out}) \delta (\phi ^p_{in}-\phi ^p_{out})
V^{\pm}_{GS}( \tilde \rho ) \eqno (II.A.2)
$$
where $\lambda$ is the string coupling constant,
$\delta (x^j_{in}-x^j_{out})
\delta (s^a_{in}-s^a_{out}) \delta (\phi ^p_{in}-\phi ^p_{out})
$ is an overlap delta function in the
string fields,
$\tilde \rho$ is the interaction point where either one string splits
or two strings join,
$$V^+
_{GS}(\tilde \rho )=\lim _{\rho \to {\tilde \rho}}[ \partial _{\rho}
x^L (\rho - \tilde \rho)^{1 \over 2} + \epsilon_
{klmn}\partial_{\rho}
 x^{kl}
s^{+m} s^{+n} (\rho -\tilde \rho )^{3 \over 2}
+ \partial _{\rho} x^R \epsilon _{klmn} s^{+k}
s^{+l}  s^ {+m} s^{+n} (\rho - \tilde \rho )^{5 \over 2}]$$
is the interaction-point operator
if $\hat s^{-\bar l}
 \equiv \sqrt {\dz\rho}\, s^{-\bar l}$ is chosen
to go like $\sqrt {z-\tilde z}$ near the interaction point for
$l$=1 to 4,
$$V^-
_{GS}(\tilde \rho )=\lim _{\rho \to {\tilde \rho}}[ \partial _{\rho}
x^R (\rho - \tilde \rho)^{1 \over 2} + \partial_{\rho} x^{kl}
s^{-\bar k} s^{-\bar l} (\rho -\tilde \rho )^{3 \over 2}
+ \partial _{\rho} x^L \epsilon ^{klmn} s^{-\bar k}
s^{-\bar l} s^{-\bar m}  s^{-\bar n}
(\rho - \tilde \rho )^{5 \over 2}]$$
is the interaction-point operator
if $\hat s^{+l} \equiv \sqrt {\dz\rho }\, s^{+l}$ is chosen
to go like $\sqrt {z-\tilde z}$ near the interaction point for
$l$=1 to 4,
and $x^L\equiv
x^7+ix^8$, $x^R\equiv x^7-ix^8$, $x^{kl}\equiv
\sum_{j=1}^6 \sigma ^{kl}_j
x^j $ with $\sigma ^{kl}_j$ the Clebsch-Gordan coefficients for
combining two SO(6) spinors into an SO(6) vector. Note that because
of the singular behavior of $\partial _\rho x^j$ and $s^a$ near the
interaction point, $\tilde \rho$, the limiting procedure in
$V_{GS}^{\pm}$ is well-defined.

The easiest way to show that this Green-Schwarz light-cone
interaction term leads to Lorentz-covariant scattering
amplitudes is by comparing with the light-cone interaction
term in the Neveu-Schwarz-Ramond formalism where Lorentz
invariance has already been proven.$^{20}$ When using an ordinary
worldsheet, the Neveu-Schwarz-Ramond light-cone interaction
term is:$^{12}$
$$L_{NSR}^{int}=\lambda \delta (x^j_{in}-x^j_{out})
\delta (\Gamma^j_{in}-\Gamma^j_{out})
V_{NSR}( \tilde \rho ) \eqno (II.A.3)
$$
where $\Gamma ^j$ is a fermionic SO(8) vector, and
$V_{NSR}(\tilde\rho)=\lim _{\rho \to \tilde \rho} [\drho x^j \Gamma^j
(\rho - \tilde \rho )^{3 \over 4}]$.

It is easy to check that the Green-Schwarz interaction-point operator,
$V_{GS}$, is related to $V_{NSR}$ through SO(8) triality since both
V's contract $\drho x^j$ with the field that creates a
vector boson of polarization j out of the ground state (in the
Green-Schwarz formalism, the ground state is either a massless
vector boson of polarization 7+i8 or 7-i8, depending on the
boundary conditions of the ($s^{+l}, s^{-\bar l}$) fields).$^3$ Since
the free part of the light-cone Green-Schwarz action is also related
through SO(8) triality to the free part of the light-cone
Neveu-Schwarz-Ramond action,$^{21}$ the two light-cone formalisms of the
superstring will give the same S-matrix scattering amplitudes.

There are two problems that occur with this light-cone method
on ordinary worldsheets, either in the Green-Schwarz or in the
Neveu-Schwarz-Ramond formalism. The first problem is that
because the interaction is not just an overlap delta function, the
integrand of the scattering amplitude will have momentum-dependent
terms that depend explicitly on the location of the interaction
points (these terms come from correlation functions involving the
interaction-point operators). Because it is extremely difficult to
solve for the locations of these interaction points in terms of
the modular parameters of the Riemann surface, only the simplest
scattering amplitudes have been evaluated using these methods.$^{2,3,4}$

A second problem is that when two interaction-point operators
collide, the integrand of the scattering amplitude becomes
divergent due to contractions between the interaction-point
operators. In order to regularize this unwanted divergence,
it is necessary to modify the light-cone interaction term
to include higher-order contact terms between four and more strings.$^{22,23}$
The exact form of these contact terms in either
the Green-Schwarz or the Neveu-Schwarz-Ramond formalism is not yet
known.

In the Neveu-Schwarz-Ramond formalism, both of these problems
were solved by putting the light-cone string fields on an
N=1 super-worldsheet, rather than on an ordinary worldsheet.$^{13}$
It was proven that integrating the exponential of the light-cone
action over all N=1 super-worldsheets with the interaction term
being a simple delta function in the string superfields is
equivalent to integrating over all ordinary worldsheets with
the complicated interaction term described in equation (II.A.3).
With this simplification of the interaction term, the integrands
of the scattering amplitudes no longer depend explicitly on
the locations of the interaction points and therefore contain
no divergences when two interaction points collide. Furthermore,
because all terms in the integrands of the scattering amplitudes
can now be easily expressed as functions of modular parameters
of N=1 super-Riemann surfaces, it is possible to explicitly
evaluate Neveu-Schwarz-Ramond multi-loop scattering amplitudes
using this new light-cone method.$^{14,15}$

The trick that was used to prove the equivalence of the new
and old light-cone methods in the Neveu-Schwarz-Ramond
formalism was to analyze the behavior of the Wick-rotated light-cone
N=1 super-worldsheet coordinates, $\rho$ and $\psi$, near
the interaction points.
By superconformally transforming the light-cone interacting
string super-worldsheet onto a smooth punctured N=1
super-Riemann surface, $\rho$ can be expressed as an analytic
function of the  super-Riemann surface's cordinates, $z$ and $\eta$
($\psi $ is determined from $\rho$ by the superconformal
condition, $\psi =-i (\dz \rho )^{-{ 1 \over 2}} D_\eta \rho $).
This means that by suitably choosing values for $\tilde z$ and
$\tilde \eta$, there always exist values for $\tilde \rho$ and
$\tilde \psi$ such that near the interaction point,
$\rho -\tilde \rho -i( \rho - \tilde\rho)^{-{ 1\over 4}} \psi
\tilde\psi \rightarrow a(z - \tilde z -i\eta\tilde\eta )^2$ and
$\psi
-( \rho - \tilde\rho)^{-{ 1\over 4}} \tilde\psi
\rightarrow \sqrt{2a} (\eta -\tilde\eta ) \sqrt{
z-\tilde z -i\eta\tilde\eta}$ where $a$ is a constant.

Since the coordinates of an ordinary worldsheet would have no $\tilde\psi$
dependence near the interaction
point, one needs to check how the free light-cone Neveu-Schwarz-Ramond
action, $I^{free}_{NSR}$, transforms under the local super-reparameterization,
$[\rho ^\prime =
\rho -i( \rho - \tilde\rho)^{-{1\over 4}} \psi
\tilde\psi$, $\psi ^\prime =
\psi
-( \rho - \tilde\rho)^{-{1\over 4}} \tilde\psi ]$. It is
straightforward to show that under this reparameterization,
${I^{free}_{NSR} }{\,}' =I^{free}_{NSR} +\tilde\psi
V_{NSR} (\tilde\rho ),$
where $V_{NSR} (\tilde\rho ) $ is defined in equation (II.A.3).
Therefore, integrating over the $\tilde\psi$'s,
as well as the $\tilde\rho$'s
and the twists, of the interacting super-worldsheet
pulls down the appropriate interaction-point operators from the
exponential of the light-cone Neveu-Schwarz-Ramond action. For
more details on this equivalence proof in the light-cone
Neveu-Schwarz-Ramond formalism, see reference 13.

In order to repeat this trick for the light-cone Green-Schwarz
formalism, it is first necessary to rewrite the Green-Schwarz
interaction-point operator, $V^{\pm}_{GS}$, in a form that more
closely resembles $V_{NSR}$. The first step is to make $V^{\pm}_{GS}$
fermionic by choosing the ground state (i.e., the state annihilated
by the zero modes of either $s^{+l}$ or $s^{-\bar l}$, for $l$=1 to 4)
to be one component of a massless spinor fermion, rather than
one component of a massless vector boson. By breaking SO(8) down
to U(4) in this way, the SO(8) vector splits into a ($4_{-{ 1 \over 2}},
\bar 4_{+{ 1\over 2}}$) representation of U(4), whereas it is the SO(8)
anti-chiral spinor that splits into a ($1_{+1}, 6_0, 1_{-1}$)
representation of U(4) (the remaining SO(8) chiral spinor still
splits into a ($4_{+{ 1\over 2}}, \bar 4_{-{ 1\over 2}}$) representation).
With this choice of ground state,
$$V^+
_{GS}(\tilde \rho )=\lim _{\rho \to {\tilde \rho}}[
\partial_{\rho}
x^{+\bar l}
s^{+l} (\rho -\tilde \rho )
+\epsilon_{klmn} \partial _{\rho} x^{-k}
s^{+l}  s^ {+m} s^{+n} (\rho - \tilde \rho )^2]$$
is the interaction-point operator
if $\hat s^{-\bar l}
$ is chosen
to go like $\sqrt {z-\tilde z}$ near the interaction point for
$l$=1 to 4,
and
$$V^-
_{GS}(\tilde \rho )=\lim _{\rho \to {\tilde \rho}}[
 \partial_{\rho} x^{-l}
s^{-\bar l} (\rho -\tilde \rho )
+\epsilon^{klmn} \partial _{\rho}  x^{+\bar k}
s^{-\bar l}  s^{-\bar m}  s^{-\bar n}
(\rho - \tilde \rho )^2]$$
is the interaction-point operator
if $\hat s^{+l} $ is chosen
to go like $\sqrt {z-\tilde z}$ near the interaction point for
$l$=1 to 4,
where $x^{-l}\equiv x^l -ix^{l+4}$ and $x^{+\bar l} \equiv x^l +i x^{l+4}$
for $l$=1 to 4.
Note that $V^{\pm}_{GS}$ is still constructed by contracting $\drho
x^j$ with the field that constructs a massless vector boson of
polarization j out of the ground state.

The second step to make $V^{\pm}_{GS}$ closer resemble $V_{NSR}$
is to integrate over light-cone worldsheets with either of the two
types of boundary conditions at the interaction points, rather than
integrating only over worldsheets with a fixed set of interaction-point
boundary conditions.
Since the term in $V_{GS}^+$ (or $V_{GS}^-$) that is cubic in
$s^{+l}$ (or
$s^{-\bar l}$) becomes linear in $s^{-\bar l}$
(or $s^{+l}$) if the boundary conditions
at the interaction point are flipped, the
Green-Schwarz interaction-point operator can be written in the
following simple form:
$$V^+_{GS}=\lim_{\rho \to \tilde\rho} [\drho  x^{+\bar l}s^{+l}
(\rho -\tilde\rho )]~~~ \hbox {    and    }~~~
V^-_{GS}=\lim_{\rho \to \tilde\rho} [\drho x^{-l} s^{-\bar l}
(\rho-\tilde\rho)],
\eqno (II.A.4)$$
where worldsheets with either type of boundary conditions at the
interaction points are allowed (note that the boundary conditions
at the ends of the external strings are still kept fixed).

It will now be shown that after combining the light-cone heterotic
Green-Schwarz fields into N=(2,0) superfields, integration over N=(2,0)
super-worldsheets
with the interaction term being a simple overlap delta function in the
string superfields is equivalent to integration over ordinary worldsheets
with the non-trivial interaction term described in equations (II.A.2) and
(II.A.4).
\vfill\eject
\centerline {\bf II.B. The Light-Cone Method on an N=(2,0)
Super-Worldsheet}

An N=(2,0) super-worldsheet is parameterized by two commuting coordinates,
$\tau$ and $\sigma$, and by two anti-commuting coordinates, $\psi^+$
and $\psi^-$.$^{24}$ In two-dimensional Minkowski space, $\psi^+$ and $\psi^-$
are complex conjugates, while in Euclidean space, they are unrelated.
The $x^j$ and $s^a$ fields combine into four pairs of chiral and
anti-chiral bosonic superfields in the following way:
$$X^{+\bar  l}(\rho^+, \pp ,\bar\rho )=x^{+
\bar l}(\rho^+ ,\bar \rho )+i\pp s^{-\bar l}(\rho^+ ,\bar\rho )
\hbox{ with }\Dm X^{+ \bar l}=0 \hbox{ for $l$=1 to 4,}$$
$$
X^{-l}(\rho^-, \pd ,\bar\rho )=x^{-
l}(\rho^- ,\bar \rho )+i\pd s^{+\bar l}(\rho^- ,\bar\rho )
\hbox{ with }\Dp X^{-l}=0 \hbox{ for $l$=1 to 4},$$
where in two-dimensional
Minkowski space, $X^{+\bar l}=(X^{-l})^*$, $\Dp\equiv \partial _{\psi^+}
+i\psi^- \drho$, $\Dm\equiv \partial_{\psi^-}+i\psi^+\drho$,
$\rho^{\pm}\equiv\tau +\sigma \pm i\psi^+\psi^-$, and $\bar \rho\equiv
\tau -\sigma$. Note that ($\psi^+$, $\psi^-$) transforms like a
($1_{+1}$, $1_{-1}$) representation under the U(4) subgroup of SO(8).

The 32 fermionic scalars, $\phi ^p$, can be similarly combined into
sixteen pairs of chiral and anti-chiral fermionic superfields in the
following way:
$$\Phi ^{+\bar q}(\rho^+, \psi^+,\bar\rho)=\phi^q(\rho^+ ,\bar\rho )+i
\phi^{q+16}(\rho^+,\bar\rho )+\psi^+t^
{+\bar q}(\rho^+, \bar\rho )\hbox { with }
\Dm\Phi^{+\bar q}=0 \hbox { for $q$=1 to 16,}$$
$$\Phi ^{-q}(\rho^-, \psi^-,\bar\rho)=\phi^q(\rho^- ,\bar\rho )-i
\phi^{q+16}(\rho^-,\bar\rho )+\psi^- t^
{-q}(\rho^-, \bar\rho )
\hbox{ with }\Dp\Phi^{-q}=0 \hbox { for $q$=1 to 16},$$
where $\Phi^{+\bar q}
=(\Phi^{-q})^*$ in two-dimensional Minkowski
space, and ($t^{+\bar q}$, $t^{-q}$) are auxiliary fields
that will vanish due to their equations of motion.

In terms of these chiral and anti-chiral superfields, the free light-cone
action of equation (I.A.1) is simply:
$$\int d\sigma d\tau d\psi^+ d\psi^- [{i\over 4}(X^{+\bar l} \drhobar X^{-l}
-X^{-l}\drhobar X^{+\bar l})+{1\over 2}\Phi^{+\bar q}\Phi^{-q}].\eqno(II.B.1)$$

One can now consider the light-cone Green-Schwarz action on an untwisted
N=(2,0) interacting string super-worldsheet (untwisted means that
chirality can be defined globally on the worldsheet)$^{25}$. This N=(2,0)
super-worldsheet is defined by giving the N=(2,0) super-conformal
transformation, $[\rho (z, \eta^+,\eta^-), \psi^+(z+i\eta^+\eta^-,\eta^+),
\psi^-(z-i\eta^+\eta^-,\eta^-)]$, that maps the Wick-rotated
super-worldsheet onto a
punctured untwisted N=(2,0) super-Riemann surface parameterized
by the coordinates $(z,\eta^+,\eta^-,\bar z)$.
Since $s^{+l}$ and $s^{-\bar l}$ are periodic everywhere on the
light-cone Green-Schwarz worldsheet, it is possible to demand
that $\psi^+$ and $\psi^-$, as well as $\rho$, are analytic functions
of $(z,\eta^+,\eta^-)$ everywhere on the punctured N=(2,0)
super-Riemann surface (the problem of how to deal with the fermionic
zero modes$^{14,15}$
will not be discussed in this paper). In addition, $\psi^+$,
$\psi^-$, and $\rho + \bar \rho$ must be single-valued on the surface,
whereas $\rho - \bar \rho$
is allowed to shift by a constant when going around the 2g non-trivial
loops of the genus g surface (these constants correspond to the usual
twists).
Finally, the condition that the transformation is N=(2,0)
superconformal and untwisted implies that $\Dep\rho =i\psi^-\Dep\psi^+$,
$\Dem\rho =i\psi^+\Dem\psi^-$, and $\Dep\psi^-=\Dem\psi^+ =0$.

As in the Neveu-Schwarz-Ramond formalism, one can measure the effect of
the super-worldsheet on the light-cone action by comparing the
local behavior of $[\rho ,\psi^+,\psi^-]$ near the interaction point
of an ordinary worldsheet
(i.e., the point where $\dz\rho=0$)
with the local behavior near the interaction
point of a general N=(2,0) super-worldsheet.
Note that away from
the interaction points, the light-cone action feels no effect from
the super-worldsheet because away from these special points,
the local behavior of $[\rho ,\psi^+,\psi^-]$ is the same for
ordinary conformal transformations as for general N=(2,0) superconformal
transformations.

If the
transformation, $[\rho (z, \eta^+,\eta^-), \psi^+(z+i\eta^+\eta^-,\eta^+),
\psi^-(z-i\eta^+\eta^-,\eta^-)]$,
were an ordinary conformal transformation, $\rho -\tilde
\rho$ would go like
$a(z-\tilde z)^2$ near the interaction point, implying
through the analyticity of $\psi^{\pm}$ that either $\psi^+$
goes like $b^{-1}a\eta^+$ and $\psi^-$ goes like $2b\eta^- (z-
\tilde z)$, or $\psi^+$ goes like $2b\eta^+(z-\tilde z)$ and
$\psi^-$ goes like $b^{-1}a\eta^-$. Since the behavior of
$\psi^+$ and $\psi^-$ is inversely correlated with the behavior
of $s^{-\bar l}$ and $s^{+l}$, the first
case corresponds to the interaction-point
boundary condition $\hat s^{-\bar l}\equiv \sqrt{\dz\rho}
\, s^{-\bar l}
\rightarrow \sqrt{z-\tilde z}$ for $l$=1 to 4, whereas the
second case corresponds to the boundary condition
$\hat s^{+l}\equiv \sqrt{\dz\rho}\, s^{+l}
\rightarrow \sqrt{z-\tilde z}$ for $l$=1 to 4.

For a general N=(2,0) superconformal transformation, there are also
two possible cases corresponding to the two different types of
interaction-point boundary conditions. In the first case,
$$\rho -\tilde\rho-i\psi^+\tilde\psi^-\rightarrow
a(z-\tilde z -i\eta^+\tilde\eta^--i\eta^-\tilde\eta^+)^2,$$
$$\psi^+\rightarrow b^{-1}a(\eta^+ -\tilde\eta^+ ),\quad\hbox{and}\quad
\psi^- -\tilde\psi^-
\rightarrow 2b(\eta^- -\tilde\eta^-)
(z-\tilde z -i\eta^+\tilde\eta^--i\eta^-\tilde\eta^+),$$ for some values
of $\tilde \rho$, $\tilde\psi^-$, and $(\tilde z,\tilde \eta^+,
\tilde\eta^- )$. In the second case,
$$\rho -\tilde\rho-i\psi^-\tilde\psi^+\rightarrow
a(z-\tilde z -i\eta^+\tilde\eta^--i\eta^-\tilde\eta^+)^2,$$
$$\psi^+ -\tilde\psi^+
\rightarrow 2b(\eta^+ -\tilde\eta^+)
(z-\tilde z -i\eta^+\tilde\eta^--i\eta^-\tilde\eta^+),\quad\hbox{and}\quad
\psi^- -\tilde\psi^-\rightarrow b^{-1}a(\eta^- -\tilde\eta^- ),$$
for some values
of $\tilde \rho$, $\tilde\psi^+$, and $(\tilde z,\tilde \eta^+,
\tilde\eta^- )$.

Suppose that the first type of boundary conditions (i.e.,
$\hat s^{-\bar l} \rightarrow \sqrt{z-\tilde z}$ for $l$=1 to 4) is
imposed near the interaction point.
One then has to measure the change of the light-cone action
under the super-reparameterization $[\rho^\prime=\rho-i\psi^+
\tilde\psi^-, {\psi^+}{}'=\psi^+, {\psi^-}{}'=\psi^- -
\tilde\psi^-]$ near the interaction point. Naively, the action
would not change under this super-reparameterization because the
action is invariant under N=(2,0) superconformal transformations.
However, the interaction-point boundary conditions of $\psi^+$
and $\psi^-$, and therefore of $s^{-l}$ and $s^{+\bar l}$, are not
preserved under this super-reparameterization ($\psi^-{}'$ no longer
goes like $\eta^-(z-\tilde z)$). The easiest way to measure the
effect of this change of boundary conditions on the light-cone
action is to define the action locally in terms of fields which
do not require any boundary conditions, namely by locally
replacing $s^{-\bar l}$ and $s^{+l}$ with $\check s^{-\bar l}\equiv
(\rho -\tilde \rho )^{-{1\over 4}}s^{-\bar l}$ and $\check s^{+l}\equiv
(\rho -\tilde \rho )^{ 1\over 4}s^{+l}$ ($\check s^{-\bar l}$ and
$\check s^{+l}$
have the same behavior near the interaction point as the
Neveu-Schwarz-Ramond fermionic field, $\Gamma^j$). Conversely,
one can locally replace the super-worldsheet coordinates with
the coordinates, $\check \psi^+\equiv(\rho -\tilde\rho )^{1 \over 4}\psi^+$
and $\check \psi^-\equiv(\rho -\tilde\rho )^{-{1\over 4}}\psi^-$. In
terms of these new coordinates, the super-reparameterization
is $[\rho^\prime=\rho -i(\rho -\tilde\rho)^{-{1\over 4}}\check\psi^+
\tilde\psi^-, \check \psi^+{}'
=\check\psi^+,\check\psi^-{}'=
\check\psi^- -
(\rho -\tilde\rho )^{-{ 1\over 4}}\tilde\psi^-]$,
and the change in the free light-cone action, $I_{GS}^{free}$, near the
interaction point is calculated as follows:
$$I_{GS}^{free}{}'={1\over 4}\int d\sigma d\tau d\psi^+ d\psi^-
[i(X^{+\bar l}(\rho^+{}',\psi^+ , \bar \rho ) \drhobar
 X^{-l}(\rho^- ,\psi^-{}',\bar\rho) $$
$$ - X^{-l}
 (\rho^- ,\psi^-{}',\bar\rho) \drhobar
X^{+\bar l}(\rho^+{}',\psi^+ , \bar \rho ) )+2\Phi^{+\bar q}
(\rho^+{}',\psi^+, \bar\rho)\Phi^{-q}(\rho^-,\psi^-{}',
\bar\rho )]$$
$$=
{1\over 4}\int d\sigma d\tau d\psi^+ d\psi^-
\{ e^{\gamma Q_{\check\psi^-}}
[i(X^{+\bar l} \drhobar X^{-l}
-X^{-l}\drhobar X^{+\bar l})+2\Phi^{+\bar q}\Phi^{-q}]$$
$$-i\drhobar \gamma
[X^{+\bar l}Q_{\check\psi^-}X^{-l}-X^{-l}Q_{\check\psi^-}X^{+\bar l}]\}
$$
$$=I_{GS}^{free} +\lim_{\rho\to\tilde\rho}
[{{i\pi}\over 2} \tilde\psi^-(\rho-\tilde\rho)^{3\over 4}\drho x
^{+\bar l}\check s^{+l}]
$$
where $\gamma\equiv
{-\tilde\psi^-}(\rho-\tilde\rho)^{-{1\over 4}}$,
$Q_{\check\psi^-}\equiv\partial_{\check\psi^-}-i\check\psi^+\drho$,
and the formula $$\int d\tau d\sigma f(\rho)(\rho-\tilde\rho)^{h-1}
\drhobar (\rho-\tilde\rho)^{-h}=2\pi f(\tilde\rho)~ \hbox{ for $h>0$ has
been used.}$$

Rewriting $\check s^{+\bar l}$ in terms of $s^{+\bar l}$, one sees
that under the super-reparameterization,
$[\rho^\prime=\rho-i\psi^+
\tilde\psi^-$, ${\psi^+}{}'=\psi^+, {\psi^-}{}'=\psi^- -
\tilde\psi^-]$ near the interaction point, the free light-cone
action picks up an interaction term equal to  ${i\pi\over 2}
\tilde\psi^- V^+_{GS}$, where $V^+_{GS}$ is defined in equation
(II.A.4).

If, on the other hand, the second type of
interaction-point boundary conditions
are imposed (i.e., $\hat s^{+l}$ goes like $\sqrt{z-\tilde z}$
for $l$=1 to 4 near the interaction point), the same reasoning
can be used to show that the free light-cone action picks
up an interaction term equal to
$\lim_{\rho\to\tilde\rho}[{i\pi\over 2}\tilde\psi^+
(\rho-\tilde\rho)^{3\over 4}\drho
x^{-l} \check s^{-\bar l}]$=
${i\pi\over 2} \tilde\psi^+
V^-_{GS}$.

So by integrating
over either $\tilde\psi^-$ or $\tilde\psi^+$ for each interaction
point of the super-worldsheet, the appropriate interaction-point
operators are pulled down from the exponential of the
light-cone action. It has therefore been proven that integrating
over N=(2,0) super-worldsheets the exponential of the ``free''
light-cone action (``free'' means that the interaction term is
a simple overlap delta function in the superfields) is
equivalent to integrating over ordinary worldsheets the
exponential of the light-cone action with the non-trivial
interaction term described in equations (II.A.2) and (II.A.4).
Since the integrands of the scattering amplitudes in this new
Green-Schwarz light-cone method no longer have explicit
dependence on the location of the interaction points, the
problem of divergences coming from the collision of interaction
points does not occur (it has
been assumed that the measure factor contains no singularities
when two interaction points collide), and the explicit calculation
of superstring scattering amplitudes in the light-cone gauge
should simplify.

The light-cone moduli for a genus g super-worldsheet with
n external strings consists of 2g-2+n fermionic moduli (there is
only one $\tilde\psi$ for each interaction point) and 6g-6+2n
bosonic moduli (the 2g-2+n $\tilde\rho$'s and
$\tilde{\bar\rho}$'s, the 2g
twists in the $\sigma$ direction,  minus 2 constants due to
energy conservation and rotational invariance). In addition, one
has to specify at each interaction point which type of
boundary condition has been imposed. The $2^{2g-2+n}$ different
types of boundary conditions that must be summed over is
reminiscent of the $2^{2g}$ different types of spin structures
that must be summed over in the Neveu-Schwarz-Ramond
superstring formalism.

Because a generic punctured N=(2,0) super-Riemann surface has
more moduli than the light-cone N=(2,0) super-worldsheets, the
light-cone super-worldsheets correspond to a restricted class
of punctured N=(2,0) super-Riemann surfaces. The restriction
is that there exists on the punctured super-Riemann surface
a real single-valued bosonic superfield, T, and two
complex single-valued fermionic superfields, $\Psi^+$ and $\Psi^-$
(in two-dimensional Minkowski space, $(\Psi^+)^*=\Psi^-$),
satisfying:
$$\dzbar\Psi^+=\dzbar\Psi^-=
\Dem\Psi^+=\Dep\Psi^-=0, \eqno(II.B.2)$$
$$\hbox{and}\quad
\dz T=\Dep\Psi^+\Dem\Psi^-
-i\Psi^+\dz\Psi^- -i\Psi^-\dz\Psi^+,$$
where $(z,\eta^+,\eta^-,\bar z )$ are coordinates for
the N=(2,0) super-Riemann surface (in two-dimensional
Euclidean space, $\Psi^+$ and $\Psi^-$ are
unrelated and the condition on T implies
only that $Re[\int_C dz d\eta^+
d\eta^- (\Psi^+\Psi^- )]=0$ around any closed loop C). With this
restriction, the punctured N=(2,0) super-Riemann surfaces have the
same number of moduli as the light-cone super-worldsheet (T and
$\Psi^{\pm}$ are just the $\rho +\bar\rho$ and $\psi^{\pm}$
coordinates of the super-worldsheet).

It will now be shown how the heterotic Green-Schwarz scattering
amplitudes defined as integrals over this restricted class of
N=(2,0) super-Riemann surfaces can be derived starting from a
manifestly Lorentz-covariant approach.

\vskip 24pt
\centerline {\bf III.A. The Covariant Method on an Ordinary Worldsheet}

In 1984, Green and Schwarz discovered a Lorentz-covariant action on
an ordinary worldsheet for the Green-Schwarz superstring that had
enough classical symmetries to be able to be gauge-fixed to the
free light-cone action.$^6$
For the ten-dimensional heterotic Green-Schwarz superstring, this
action is:
$$\int d\tau d\sigma \{ {1 \over det~e}[\pi_-^\mu \pi_+^\mu
+e_-^m \phi^p\partial_m \phi^p ]
+i
\partial_{\bar\rho} x^{\mu} (\partial_\rho \tgt)
-i\partial_\rho x^{\mu} (\partial_{\bar \rho} \tgt)\} , \eqno(III.A.1)$$
where $e_{\pm}^m$ is the two-dimensional vielbein
($m$ is either $\tau$ or $\sigma$), $\gamma^{\mu}_{\alpha\beta}$
are the symmetric SO(9,1) gamma-matrices satisfying the cyclic identity
$\gamma^{\mu}_{\alpha\beta}\gamma_{\mu~\gamma\delta}+
\gamma^{\mu}_{\beta\gamma}\gamma_{\mu~\alpha\delta}+
\gamma^{\mu}_{\gamma\alpha}\gamma_{\mu~\beta\delta}=0$,
$\pi_{\pm}^\mu\equiv e_{\pm}^m(\pim)$, and in
two-dimensional Minkowski space, $\rho\equiv\tau+\sigma$ and
$\bar\rho\equiv\tau-\sigma$ are the worldsheet
coordinates, + and $-$ are the
tangent-space coordinates, $x^{\mu}$ is a real SO(9,1) vector,
$\theta^{\alpha}$ is a real SO(9,1) chiral spinor, and $\phi^p$
are the 32 real SO(9,1) scalars that parameterize the self-dual
lattice of the heterotic string.

As in the light-cone gauge, $\theta^{\alpha}$ must be periodic everywhere
on the surface in order to preserve the global spacetime supersymmetry
transformation,
[$\delta\theta^\alpha=\epsilon^\alpha$, $\delta x^\mu =i\theta^\alpha
\gamma^{\mu}_{\alpha\beta}\epsilon^\beta$], whereas $\phi^p$ can
be either periodic or anti-periodic.

This action is classically invariant under the usual worldsheet
reparameterizations, Weyl scaling transformations, and two-dimensional
Lorentz rotations, as well as under the following
$\kappa_{\alpha}$-transformations:$^{26}$
$$[\delta\theta^\alpha=(\gamma_{\mu}^{\alpha\beta}\kappa_{\beta})
\pi_-^\mu,~\delta x^\mu=i(\delta \tgt), ~\delta e_+^m=
4i(e_+^n \partial_n \theta^\alpha \kappa_\alpha)e_-^m]$$
where $\kappa_\alpha$ is a real SO(9,1) anti-chiral spinor.

In order to write this action in terms of free fields, it is
necessary to use the $\kappa_\alpha$ transformations to gauge-fix
to zero the eight components of $\gamma^+_{\alpha\beta}\theta^\beta$
(since $\pi_-^\mu (\gamma^{\alpha\beta}_\mu \delta\theta^\beta)
=0$ on-shell, only eight of the sixteen components of $\theta^\alpha$
can be gauged away). Note that because there are no derivatives of
$\kappa_\alpha$ in the gauge transformations of $\theta^\alpha$,
this can be done everywhere without any global obstructions
(although it is true that the gauge transformation acts singularly
on $\gamma^-_{\alpha\beta}\theta^{\beta}$ when $\pi_-^+=0$,
this is not a problem since the Jacobian of the transformation is
non-singular).$^8$ In addition, the reparameterizations and
two-dimensional Lorentz rotations must be used to gauge away
all of the non-Liouville parts of the vielbein, except for the
6g-6+2n moduli of the genus g Riemann surface with n punctures.
Although this gauge choice breaks the manifest spacetime SO(9,1)
invariance down to S0(8), it preserves the classical worldsheet
conformal invariance.

The heterotic Green-Schwarz action in this semi-light-cone gauge
takes the following simple form:
$$\int d\tau d\sigma[\drho x^\mu \drhobar x_\mu +is^a \drhobar s^a +i
\phi^p \drho\phi^p]\eqno (III.A.2)$$
with the
Virasoro constraints, $\drho x^\mu \drho x_\mu +i s^a \drho s^a$=
$\drhobar x^\mu \drhobar x_\mu +i \phi^p\drhobar\phi^p$=0, where
the $s^a$ fields are the eight non-zero
components of
$(\gamma^-_{a\beta}\theta^{\beta})
(\drho x^+)^{-{1\over 2}}$.

Because all fields are free in this gauge, it is easy to
calculate the conformal anomaly using the standard
$\pm (6j^2-6j+1)$ formula that comes from a one-loop
analysis of the non-local terms in the effective
Liouville action (as noted
in reference 10, other methods of calculation
may shift the conformal anomaly into an anomaly
in the spacetime Lorentz invariance). In the left-moving sector,
the conformal anomaly receives a contribution of $-$26 from the
reparameterization ghosts, +10 from the $x^\mu$'s, and +16 from
the $\phi^p$'s. In the right-moving sector, however,
the conformal anomaly receives a contribution of $-$26 from the
reparameterization ghosts, +10 from the $x^\mu$'s, and +4 from
the $s^a$'s, and therefore is non-zero (note that because
the $\kappa_\alpha$-transformation does not involve derivatives of
$\kappa_\alpha$, there are no propagating ghosts coming from the
gauge-fixing of $\gamma^+_{\alpha\beta}\theta^\beta$). The presence of
a conformal anomaly means that the usual method for regularizing the
semi-light-cone gauge free-field action can not be used because it fails
to preserve the classical symmetries of the action$^{11}$
(although there
exist alternative gauges for the Green-Schwarz action on an ordinary
worldsheet in which the conformal anomaly is claimed to
vanish,$^{27}$ the Green-Schwarz
action in these gauges is not a free-field
action, making quantization and amplitude calculations impractical).

The fact that the usual regulator can not be used in the
semi-light-cone gauge is not surprising, since otherwise, one
would have a free-field action for the Green-Schwarz superstring
defined on ordinary Riemann surfaces. It is clear that such an
action would lead to non-unitary scattering amplitudes in the
Polyakov approach, since the amplitudes would contain neither
the interaction-point operators that are present in the old
light-cone method for calculating scattering amplitudes, nor
the fermionic moduli that are present in the new light-cone
method.

In the Neveu-Schwarz-Ramond formalism for the superstring, the
Polyakov approach does not have this inconsistency
because the exponential of the covariant action is integrated
over N=1 super-Riemann surfaces, rather than over ordinary Riemann
surfaces.$^7$ The obvious guess for the heterotic Green-Schwarz formalism
is therefore to integrate the exponential of the covariant
action not over ordinary Riemann surfaces, but over N=(2,0)
super-Riemann surfaces. In order to do this, one first has
to construct a Lorentz-covariant action for the heterotic
Green-Schwarz superstring on an N=(2,0) super-worldsheet.

In constructing the covariant action for the Green-Schwarz
superstring on an ordinary worldsheet, it was useful to
first study the covariant action for the superparticle on
a worldline.$^{26}$ To construct the covariant superstring action
on an N=(2,0) super-worldsheet, it turns out also to be useful
to first study the covariant superparticle action on an
N=2 super-worldline.

\vskip 24pt
\centerline {\bf III.B. The Covariant Superparticle Action on an N=2
Super-Worldline}

The superparticle on both an N=1 and N=2 super-worldline was
first discovered by Sorokin, Tkach, and Volkov in
1989.$^{28}$ Although they considered three and four-dimensional
target spaces, it is straightforward to generalize their
results to a ten-dimensional target space.$^{19,29}$ Instead of
starting from the usual action for the massless particle,
$\int d\tau [a \partial_\tau x^\mu \partial_\tau x_\mu]$, they
started from an action containing bosonic spinors that closely
resemble the twistor variables of Penrose$^{30}$. This action is:
$$\int d\tau \{ p_\mu [\partial_\tau x^\mu - (\lambda^{\alpha}\gamma^\mu_
{\alpha\beta}\lambda^{\beta})]\} , \eqno(III.B.1)$$
where $\lambda^\alpha$
is a bosonic SO(9,1) chiral spinor.

By varying
the fields, one finds $\dt x^\mu-(\lambda^{\alpha}
\gamma^{\mu}_{\alpha\beta}\lambda^\beta)$=
$\dt p^\mu$=
$p_\mu(\gmu \lambda^\beta)
=0$.
Since $p_\mu (\gmu\lambda^\beta)=0$ implies that $p^\mu=a(\lambda
^\alpha\gmu\lambda^\beta)$ for some $a$, these
equations of motion are equivalent to the usual massless
equations, $(\dt x^\mu)^2=\dt (a\dt x^\mu)=0$, except
for the fact that $\dt x^0$ is required to be non-negative
($\dt x^0=\lambda^{\alpha}\gamma^0_{\alpha\beta}\lambda^\beta$
=$\sum_{\alpha =1}^{16} (\lambda^\alpha)^2$). Note that nine components
of $p_\mu$ are auxiliary fields, while the tenth component plays
the role of a metric. The advantage of starting from this twistor-like
action is that supersymmetrizing the worldline produces the
superparticle, rather than the spinning particle.

For example, the superparticle action on an N=1 super-worldline
parameterized by $\tau$ and $\psi$ is given by:
$$\int d\tau d\psi [-i P_\mu ~\Pi_\psi^\mu]~ ,
 \eqno(III.B.2)$$
$$\hbox{where }D_\psi=\partial_\psi+i\psi\dt,\quad
\Pi_\psi^\mu= D_\psi X^\mu -i( D_\psi\TgT),$$
$$X^{\mu}=x^\mu+i\psi\Gamma^\mu,\quad
\Theta^\alpha=\theta^\alpha+\psi\lambda^\alpha,\quad \hbox{ and}\quad
P^\mu=p^\mu+i\psi q^\mu.$$

After varying the N=1 superfields, one finds
the equations of motion to be $\Pi_\psi^\mu$=$D_\psi P^\mu$=
$P_\mu (\gmu D_\psi\Theta^\beta)$=0, implying that $P^\mu$=$A (D_\psi\Theta
^\alpha\gmu D_\psi\Theta^\beta)$ for some superfield $A=a+i\psi b$.
Since $D_\psi\Theta^\alpha\gmu
D_\psi\Theta^\beta$=$\dt X^\mu-i(\dt\tgt)$
$\equiv \Pi_\tau^\mu$, one finds $(\Pi^\mu_\tau)^2$=
$D_\psi[A\Pi_\tau^\mu]$=0, which implies in component form the
usual superparticle equations of motion, $(\pi^\mu_\tau)^2$
=$\dt(a \pi_\tau^\mu)$=   $\pi_{\tau\,\mu}(\gmu\dt\theta^\beta)$=0,
where $\pi_\tau^\mu\equiv\dt x^\mu-i(\dt\tgt)$.
The remaining equations of motion fix the values of the
auxiliary fields ${1\over a}p^\mu$, $q^\mu$, $\Gamma^\mu$, and nine
components of $\lambda^\alpha$ (the other seven components of
$\lambda^\alpha$ are gauge fields).

It is easy to show that the
superparticle action of equation (III.B.2) is invariant
under the
N=1 super-reparameterizations,
$$[\delta\tau=
2R-\psi D_\psi R,~
\delta\psi=-iD_\psi R]$$ where R is a bosonic superfield (under
this super-reparameterization, $\delta D_\psi=(\partial_\tau R) D_\psi$),
and under the seven independent $K_{\beta}$-transformations,$^{31}$
$$[\delta\Theta^\alpha=(D_\psi\Theta^\gamma \gamma_{\gamma\delta}^\mu
D_\psi\Theta^\delta)(\gamma_\mu^{\alpha\beta}K_\beta)-2D_\psi
\Theta^{\alpha}(D_\psi\Theta^\beta K_\beta),~ \delta X^\mu=
i(\delta\Theta^{\alpha}\gmu\Theta^\beta)]$$
where $K_\beta$ is a fermionic SO(9,1) anti-chiral spinor
superfield (there are only seven, rather than sixteen independent
transformations since $D_\psi\Theta^\alpha\gmu\delta\Theta^\beta$
vanishes identically). These symmetries are enough to get to
light-cone gauge since the fermionic part of the N=1
super-reparameterizations replaces the
missing eighth $\kappa_{\beta}$-transformation (this differs from
(supersymmetric$)^2$ systems$^{32}$ in which the
super-reparameterization invariance
is an additional symmetry), and the seven
bosonic parts of the $K_{\beta}$-transformations can be used
to gauge-fix the remaining $\lambda^\alpha$ fields.
Of course, the action is also invariant under the global
spacetime supersymmetry transformation, [$\delta\Theta^\alpha=
\epsilon^\alpha$, $\delta X^\mu=i(\Theta^\alpha\gmu\epsilon^\beta)].$

Amazingly, the superparticle action on an N=2 super-worldline can
also be constructed by supersymmetrizing the twistor-like action
of equation (III.B.1).$^{28}$ On an N=2 super-worldline parameterized
by $\tau$, $\psi^+$, and $\psi^-$, the superparticle action is:
$$\int d\tau d\psi^+ d\psi^- [-i(P_\mu \Pi_\pp^\mu - \bar P_\mu
\Pi_\pd^\mu)] \eqno (III.B.3),$$
$$\hbox{where }D_{\psi^\pm}=\partial_{\psi^\pm} +i\psi^\mp\dt,
 \quad\Pi_{\psi^\pm}^\mu=D_{\psi^\pm}X^\mu-i(D_{\psi^\pm}\TgT),$$
$$P^\mu=p^\mu+\pp q^\mu +\pd r^\mu +i\pp\pd s^\mu,~\quad
\bar P^\mu=\bar p^\mu+\pd \bar q^\mu +\pp \bar r^\mu +i\pd\pp
\bar s^\mu,$$
$$X^\mu=x^\mu+i\pp\Gamma^\mu+i \pd\bar\Gamma^\mu
+\pp\pd h^\mu, \quad~\Theta^\alpha=\theta^\alpha+\pp\lambda^\alpha
+\pd\bar\lambda^\alpha+\pp\pd f^\alpha,$$
$$\hbox{and}\quad (\pp)^*=\pd,~
(\Dp)^*=\Dm,~ (P^\mu)^*=\bar P^\mu,~ (X^\mu)^*=X^\mu,~
(\Theta^\alpha)^*=\Theta^\alpha.$$

The equations of motion for the N=2 superfields are
$$\Pi^\mu_{\psi^\pm}=
Im[\Dp P^\mu]=Im[P_\mu (\gmu\Dp\Theta^\beta)]=0.$$ Since $(\Dp)^2$=0,
$\Dp\Theta^\alpha\gmu\Dp\Theta^\beta$=0 (i.e., $\Dp\Theta^\alpha$ is
a pure spinor), implying that $P_\mu\gmu\Dp\Theta^\beta$=
$\bar P_\mu\gmu\Dm\Theta^\beta$=0 (a pure spinor with no imaginary
part must vanish). So $\Dp P_\mu(\gmu\Dp\Theta^\beta)$=0 and
$\Dp P_\mu(\gmu\Dm\Theta^\beta)$=$\Dm \bar P_\mu(\gmu\Dm\Theta^\beta)$=0.
Therefore, $$\Dp P^\mu=\Dm \bar P^\mu=A(\Dp\Theta^\alpha\gmu
\Dm\Theta^\beta)$$ for some real superfield $A=a+i\pp b+i\pd\bar b+
\pp\pd c$. Since $\Dp\Theta^\alpha\gmu\Dm\Theta^\beta$=$\Pi_\tau^\mu$,
one finds $(\Pi_\tau^\mu)^2$=$D_{\psi^\pm}[A\Pi_\tau^\mu]=0$.
In addition to implying the usual superparticle equations of motion
for $x^\mu$ and $\theta^\alpha$, these equations fix the values of
the auxiliary fields $\Gamma^\mu$, $\bar\Gamma^\mu$, $h^\mu$,
${1\over a}q^\mu$, ${1\over a}\bar q^\mu$, $s^\mu$, $\bar s^\mu$,
nineteen of the thirty-two real components of $\lambda^\alpha$,
and five complex components of $p^\mu$ and $r^\mu$ (the remaining
component fields besides $x^\mu$ and $\theta^\alpha$ are all
gauge fields).

By transforming $P^\mu$ and $\bar P^\mu$ appropriately, the action of
equation (III.B.3) is invariant under the N=2 super-reparameterizations,
$$[\delta\tau=2R-\pp\Dp R -\pd\Dm R,~ \delta\psi^\pm=-iD_{\psi^\mp} R]$$
where R is a real N=2 superfield
(under this super-reparameterization, $\delta\Dp$=  $-i(\Dp\Dm R)\Dp$
and $\delta\Dm=-i(\Dm\Dp R)\Dm$), under the six independent
$K_\beta$-transformations, $$[\delta\Theta^\alpha=(\Dp\Theta^\gamma
\gamma^\mu_{\gamma\delta}\Dm\Theta^\delta)(\gmu K_\beta)-2
\Dp\Theta^\alpha (\Dm\Theta^\beta K_\beta)-2\Dm\Theta^\alpha
(\Dp\Theta^\beta K_\beta),$$ $\delta X^\mu=i(\delta\Theta^\alpha
\gmu\Theta^\beta)]$ where $K_\beta$ is a real fermionic SO(9,1)
anti-chiral spinor N=2 superfield (only six of the sixteen transformations
are independent since $D_{\psi^\pm}\Theta^\alpha\gmu\delta\Theta^\beta$
vanishes on-shell where
$D_{\psi^\pm}\Theta^\alpha$ is a pure spinor), and under
the five independent chiral $C^\alpha$-transformations, $$[\delta\Theta
^\alpha=\delta X^\mu=0, ~\delta P^\mu=\Dp C^\alpha\gmu\Dp\Theta^\beta,~
\delta\bar P^\mu=\Dm \bar C^\alpha\gmu\Dm\Theta^\beta]$$ where
$C^\alpha$ is a complex bosonic SO(9,1) chiral spinor N=2
superfield (since $\delta P_\mu (\gmu\Dp\Theta^\beta)$ and $\Dp
\delta P_\mu$ vanish on-shell, there are only five independent
chiral complex transformations).
As in the N=1 action, these symmetries are enough to get to
light-cone gauge since the two fermionic parts of the N=2
super-reparameterizations replace the missing seventh and eighth
$\kappa_\beta$ transformations, the extra bosonic part of the
N=2 super-reparameterizations and the twelve bosonic parts of the
$K_\beta$-transformations can be used to gauge-fix the remaining
$\lambda^\alpha$ fields, the extra six fermionic parts of the
$K_\beta$-transformations can be used to gauge-fix the remaining
$f^\alpha$ fields, and the five chiral complex $C^\alpha$-transformations
can be used to gauge-fix the remaining $p^\mu$ and $r^\mu$ fields.

Once the superparticle action on an N=2 super-worldline has
been constructed, it is straightforward to generalize to the
heterotic Green-Schwarz superstring on an N=(2,0) super-worldsheet
(although the superparticle action on an N=1 super-worldline
can also be generalized to a heterotic Green-Schwarz superstring action
on an N=(1,0) super-worldsheet,$^{33}$ this action does not allow the
correct periodicity conditions for the Green-Schwarz fields, and
it suffers from a conformal anomaly like the action on an
ordinary worldsheet).

\vskip 24pt
\centerline {\bf III.C. The Covariant Method on an N=(2,0) Super-Worldsheet}

Like the action on an ordinary worldsheet, the heterotic Green-Schwarz
superstring action on an N=(2,0) super-worldsheet contains a
superparticle-like action, the fermions from the self-dual lattice of the
heterotic string, and a Wess-Zumino term$^{34}$. Since the one-dimensional
super-vielbein of
the superparticle action is already contained in $P^\mu$ and $\bar P^\mu$,
only one component of the two-dimensional super-vielbein
needs to be introduced as an additional field in the superstring action
(alternatively, one could introduce the full two-dimensional
super-vielbein and use the gauge invariances and N=(2,0) torsion
constraints to fix all but one component).
This heterotic Green-Schwarz superstring action on an N=(2,0)
super-worldsheet parameterized by the two-dimensional Minkowski
space coordinates, $\rho\equiv\tau+\sigma$, $\bar\rho\equiv\tau-
\sigma$, $\pp$, and $\pd\equiv(\pp)^*$, is:$^{18,33,35}$
$$\int d\tau d\sigma d\pp d\pd \{
-i(P_\mu \hat\Pi_\pp^\mu - \bar P_\mu \hat\Pi_\pd^\mu) +{1\over 2}
\hat\Phi^{+\bar q}\hat\Phi^{-q} \eqno(III.C.1)$$
$$-{1\over 2}\pp[\drhobar X_\mu (\Dp\TgT)-\Dp X_\mu (\drhobar \TgT)]
$$
$$
+{1\over 2}\pd
[\drhobar X_\mu (\Dm\TgT)-\Dm X_\mu (\drhobar \TgT)]\}$$
$$\hbox{with the chirality constraints, }\quad\hDm\hat\Phi^{+\bar q}=
\hDp\hat\Phi^{-q}=0,$$
$$\hbox{and the covariant derivatives, }\quad
\hat D_{\psi^\pm}\equiv\partial_{\psi^\pm}+i\psi^\mp[\drho
+e(\rho,\bar\rho)\drhobar +(\drhobar e(\rho,\bar\rho))M],$$
where $e(\rho,\bar \rho)$ is a real component field
independent of $\psi^\pm$ and is the only remnant of the two-dimensional
super-vielbein ($\hdrho\equiv -{i\over 2}
\{ \hDp,\hDm\}$=$\drho+e\drhobar+(\drhobar e)M$),
$M$ is the generator of two-dimensional
Lorentz rotations that measures the conformal weight with respect to
$\drhobar$ (i.e., $M$ commutes with everything except for
$[M,\drhobar]=\drhobar$ and $[M,\hat\Phi]={1\over 2}\hat\Phi$),
$\hat\Pi_{\psi^\pm}^\mu\equiv\hat D_{\psi^\pm}X^\mu-i(
\hat D_{\psi^\pm}\TgT)$,
the N=2 superfields $X^\mu$, $\Theta^\alpha$, $P^\mu$, and $\bar P^\mu$,
are defined as in equation (III.B.3), and $\hat\Phi$ is defined like
$\Phi$ in equation (II.B.1), but with the appropriate modifications
due to $e$ in the chirality constraint. Note that because the
Wess-Zumino term multiplying $\pp$ (or $\pd$) is chiral (or anti-chiral)
when $\hat\Pi^\mu_{\psi^\pm}=0$, the action will be super-reparameterization
invariant after shifting $P^\mu$ and $\bar P^\mu$ appropriately.

The equations of motion one gets from varying the unconstrained
superfields are:
$$\hDp\hat\Phi^{+\bar q}=\hDm\hat\Phi^{-q}=
\hat\Pi_{\psi^\pm}=\pp\pd Re[\hDp(P_\mu \Pi^\mu_{\bar\rho})-
{i\over 2}\hat\Phi^{+\bar q}\drhobar\hat\Phi^{-q}]$$
$$=
Im[\hDp P^\mu+i\pp(\hDp\Theta^\alpha\gmu\drhobar\Theta^\beta)]
=Im[(P_\mu+{1\over 2}\pp\Pi_{\bar\rho\,\mu})(\gmu\hDp\Theta^\beta)]=0.$$
Using the same reasoning as for the superparticle, these equations
imply that
$$\hDp P^\mu
+{1\over 2}\Pi^\mu_{\bar\rho}+i\pp(\hDp\Theta^\alpha\gmu\drhobar\Theta^\beta)
=
\hDm\bar  P^\mu
+{1\over 2}
\Pi^\mu_{\bar\rho}+i\pd(\hDm\Theta^\alpha\gmu\drhobar\Theta^\beta)$$
$$=A(\hDp\Theta^\alpha\gmu\hDm\Theta^\beta)
$$
for some real N=2 superfield A, and therefore,
$$
\hat D_{\psi^\pm}(\Pi^\mu_{\bar\rho}-A
\hat\Pi^\mu_\rho)=(\hat\Pi^\mu_\rho)^2=\pp\pd[(\Pi^\mu_{\bar\rho}-
A\hat\Pi^\mu_\rho)^2+ {i\over 2}( \hat\Phi^{+\bar q}
\drhobar\hat\Phi^{-q}+\hat\Phi^{-q}\drhobar\hat\Phi^{+\bar q})]=0,$$
where $\hat\Pi^\mu_\rho\equiv\hdrho X^\mu-i\hdrho\TgT$=
$\hDp\Theta^\alpha\gmu\hDm\Theta^\beta$.
In addition to implying the usual superstring equations of motion for
the component fields, $x^\mu$, $\theta^\alpha$, and $\phi^p$,
$$\partial_-\phi^p=\partial_-\pi^\mu_+=\pi_{-\,\mu}(\gmu\partial_+
\theta^\beta)=(\pi_-^\mu)^2=(\pi_+^\mu)^2+i\phi^p\partial_+\phi^p=0,$$
where $\partial_-\equiv\drho+e\drhobar$ and $\partial_+\equiv
(1-ae)\drhobar -a\drho$, these superfield equations fix the
values of the auxiliary fields
$\Gamma^\mu$, $\bar\Gamma^\mu$, $h^\mu$,
${1\over a}q^\mu$, ${1\over a}\bar q^\mu$,
$s^\mu$, $\bar s^\mu$, $t^{+\bar q}$, $t^{-q}$,
nineteen of the thirty-two real components of $\lambda^\alpha$, and
five complex components of $p^\mu$ and $r^\mu$.

With the appropriate transformations of $P^\mu$, $\bar P^\mu$,
$\hat\Phi^{+\bar q}$, and $\hat\Phi^{-q}$, the action of
equation (III.C.1) is invariant under the N=2 super-reparameterizations,
$$[\delta\rho=2R-\pp\hDp R -\pd\hDm R,~
 \delta\psi^\pm=-i\hat D_{\psi^\mp} R,~
\delta \bar\rho=r+e\delta\rho]$$
where $R(\rho,\pp,\pd,\bar \rho)$
is a real N=2 superfield and $r(\rho,\bar\rho)$ is a real
component field independent of $\psi^\pm$
(from this super-reparameterization, $\delta\hDp =-i(\hDp\hDm R)
\hDp$
and $\delta\hDm=-i(\hDm\hDp R)\hDm$
where $\delta e=-\drho r -e\drhobar r+r \drhobar e$), under the six independent
$K_\beta$-transformations, $$[\delta\Theta^\alpha=(\hDp\Theta^\gamma
\gamma^\mu_{\gamma\delta}\hDm\Theta^\delta)(\gmu K_\beta)-2
\hDp\Theta^\alpha (\hDm\Theta^\beta K_\beta)-2\hDm\Theta^\alpha
(\hDp\Theta^\beta K_\beta),$$ $\delta X^\mu=i(\delta\Theta^\alpha
\gmu\Theta^\beta)]$,
and under
the five independent chiral $C^\alpha$-transformations, $$[\delta\Theta
^\alpha=\delta X^\mu=0, ~\delta P^\mu=\hDp C^\alpha\gmu\hDp\Theta^\beta,~
\delta\bar P^\mu=\hDm \bar C^\alpha\gmu\hDm\Theta^\beta].$$

In order to write the action in terms of free fields, it is necessary
to use the six $K_\beta$-transformations to gauge-fix to zero
$\gamma^+_{\dot a\beta}\Theta^\beta$ for $\dot a$=1 to 6, and to use the five
$C^\alpha$-transformations to gauge-fix the non-auxiliary components
of $p^\mu$, $\bar p^\mu$, $r^\mu$, and $\bar r^\mu$. Since none
of these gauge transformations involve derivatives on $K_\beta$
or $C^\alpha$, there are no propagating ghosts coming from this
gauge fixing. Furthermore, the $\pp=\pd=0$ parts of the N=(2,0)
super-reparameterizations, $R(\rho,\psi^\pm=0,\bar\rho)$ and
$r(\rho,\bar\rho)$, should be used to locally gauge-fix e and a
to zero,
giving rise to the usual right and left-moving reparameterization
ghosts of conformal weight +2.

At this point, there are two choices on how to gauge-fix the
remainder of the N=(2,0) super-reparameterizations. One choice
would be to use the remaining super-reparameterizations to
gauge-fix the remaining non-auxiliary components of $(\gamma^+_{a\beta}
\Theta^\beta)$ (this gauge choice does not require additional
propagating ghosts since the transformation of $\Theta^\alpha$
does not involve derivatives of $R$). It is easy to see
that this is just the semi-light-cone gauge that was discussed
in Section III.A. Since this semi-light-cone gauge breaks the
manifest N=(2,0) superconformal invariance of the action (only
one component of the superfield A is gauge-fixed to zero), it
is analogous in the Neveu-Schwarz-Ramond formalism of the
superstring to using the N=1 super-reparameterization invariance
to gauge-fix the light-cone component
of the fermionic SO(9,1) vector,
$\Gamma^+$, rather than using the invariance to
gauge-fix the gravitino (note that trying to regularize the
Neveu-Schwarz-Ramond action in this non-superconformal gauge
would lead to problems similar to those found for the Green-Schwarz
action in the semi-light-cone gauge).

The second choice for gauge-fixing the remainder of the N=(2,0)
super-reparameterizations is to locally gauge-fix the rest of the
superfield A to zero (this gauge choice affects the $x^\mu$
and $\theta^\alpha$ component fields through the equations of motion,
$D_{\psi^\pm}(\Pi^\mu_{\bar\rho}-A\hat\Pi^\mu_\rho)=0$).
Since A and e serve as the N=(2,0) super-vielbien, this gauge choice
requires not only the right and left-moving
fermionic ghosts of conformal weight +2, but also two right-moving
bosonic ghosts of conformal weight +${3\over 2}$, and one
right-moving fermionic ghost of conformal weight +1. Since this
gauge choice preserves the manifest N=(2,0) superconformal
invariance of the action, it is the analog of the usual
superconformal gauge in the Neveu-Schwarz-Ramond formalism.

Because six components of $\Theta^\alpha$ have been gauge-fixed
to zero, only a U(4) subgroup of the SO(9,1) Lorentz invariance
remains manifest in this N=(2,0) superconformal gauge. As
discussed in Section II.A., the SO(8) anti-chiral spinor,
$(\gamma^+\Theta)^{\dot a}$, can be chosen to break up into
a $(1_{+1}, 6_0,1_{-1})$ representation of U(4), in which case
the SO(8) chiral spinor, $(\gamma^-\Theta)^a$, breaks up into a
$(4_{+{1\over 2}},\bar 4{-{1\over 2}})$ representation of U(4),
and the SO(9,1) vector, $X^\mu$, breaks up into a $(1_0,1_0,4_{-{1\over 2}},
\bar 4_{+{1\over 2}})$ representation of U(4).

Since the constraint $\Dp\Theta^\alpha\gmu\Dp\Theta^\beta$=0
implies that $(\gamma^+\Dp\Theta)^{\dot a}(\gamma^+\Dp\Theta)^{\dot a}=0$,
it can be assumed that $\Dp[(\gamma^+\Theta)^{7}
-i(\gamma^+\Theta)^{ 8}]=0$
without loss of generality (if $
\Dp[(\gamma^+\Theta)^{ 7}
+i(\gamma^+\Theta)^{ 8}]=0$, simply exchange $\pp$
with $\pd$ everywhere). After making this choice, the constraints
$\Pi^\mu_{\psi^\pm}$=0 can be used to combine the $X^\mu$ and $\Theta^\alpha$
real superfields into the following chiral and anti-chiral complex
superfields:
$$\Psi^\pm\equiv (\gamma^+\Theta)^{ 7}
\pm i(\gamma^+\Theta)^{8},\quad
S^{+l}\equiv (\gamma^-\Theta)^l+i(\gamma^-\Theta)^{l+4},\quad
S^{-\bar l}\equiv (\gamma^-\Theta)^l-i(\gamma^-\Theta)^{l+4},$$
$$X^{+\bar l}\equiv X^l+iX^{l+4}+i \Psi^+ S^{-\bar l},
\quad X^{-l}\equiv X^l-iX^{l+4}+i\Psi^- S^{+l},
\quad X^\pm\equiv X^0\pm X^9,$$
$$\hbox{where  }\Dm\Psi^+=\Dm S^{+l}=\Dm X^{+\bar l}=\Dp\Psi^-=
\Dp S^{-\bar l}=\Dp X^{-l}=0,$$
$$(\Psi^+)^*=\Psi^-,~ (S^{+l})^*=S^{-\bar l},~
(X^{+\bar l})^*=X^{-l},~(X^+)^*=X^+, ~(X^-)^*=X^-,$$
$$\hbox{and $(\Psi^+,\Psi^-,S^{+l},S^{-\bar l},X^{+\bar l},X^{-l},X^+,X^-) $
transforms like a}$$
$$\hbox{ $(1_{+1},1_{-1},4_{+{1\over 2}},\bar 4_{-{1\over 2}},
\bar 4_{+{1\over 2}},4_{-{1\over 2}},1_0,1_0)$ representation of U(4)
for $l$=1 to 4.}$$

In terms of these complex superfields, the action of equation (III.C.1)
in N=(2,0) superconformal gauge takes the following simple form:
$$\int d\tau d\sigma d\pp d\pd [{i\over 4}
(X^{+\bar l}\drhobar X^{-l}-X^{-l}\drhobar X^{+\bar l}) +
W^-\drhobar\Psi^+ -W^+\drhobar\Psi^- +{1\over 2}\Phi^{+\bar q}
\Phi^{-q}]\eqno (III.C.2)$$
$$\hbox{with the constraints }\pp\pd[
\drhobar X^{+\bar l}\drhobar X^{-l}-
\drhobar X^-\drhobar X^++{i\over 2}(\Phi^{+\bar q}\drhobar\Phi^{-q}+
\Phi^{-q}\drhobar\Phi^{+\bar q})]=$$
$$\Dp X^+-i\Psi^-\Dp\Psi^+=\Dm X^+-i\Psi^+\Dm\Psi^-=$$
$$\Dp X^--iS^{-\bar l}\Dp S^{+l}=
\Dm X^--iS^{+l}\Dm S^{-\bar l}=$$
$$\Dp X^{+\bar l}-2iS^{-\bar l}\Dp\Psi^+=\Dm X^{+\bar l}
=\Dp X^{-l}=\Dm X^{-l}-2iS^{+l}\Dm\Psi-=
$$
$$\Dm\Psi^+=
\Dm S^{+l}=\Dm\Phi^{+\bar q}=\Dp\Psi^-=
\Dp S^{-\bar l}=\Dp\Phi^{-q}=0$$
$$\hbox{where }W^-\equiv \Psi^-(X^-+iS^{+l}S^{-\bar l})\quad \hbox{and}
\quad W^+\equiv \Psi^+(X^--iS^{+l}S^{-\bar l}).$$

It is easy
to check that the only effect of the right-moving constraints
on the superfields in the action is to fix their chiralities through
$$\Dm X^{+\bar l}=\Dp X^{-l}=\Dm\Psi^+=\Dp\Psi^-
=\Dm\Phi^{+\bar q}=\Dp
\Phi^{-q}=0;\eqno(III.C.3a)$$
to relate $W^+$ and $W^-$
through the condition
$$\Dp W^+ \Dm\Psi^- -\Dm W^- \Dp\Psi^++{
i\over 2}\Dp X^{+\bar l}\Dm X^{-l}=0;
\eqno(III.C.3b)$$
and to require that $$\Dp\Psi^+\Dm\Psi^--i\Psi^+\drho\Psi^-
-i\Psi^-\drho\Psi^+=\drho T
\hbox{ for some real superfield $T$}.\eqno(III.C.3c)$$ Note that
the $\psi^+=\psi^-=0$ component of
$\drhobar T$ should equal the $\drhobar x^+$ component
field that appears in the
left-moving Virasoro constraint.

Using the usual free-field commutation relations for the superfields
in the action, it is easy to show that the constraint in equation
(III.C.3b) generates an N=(2,0) super-Virasoro algebra$^{24}$ (note that the
requirement in equation (III.C.3c) is a global condition
and therefore does not affect the
free-field commutation relations). The $\pp\pd$
component of the constraint is just the usual right-moving Virasoro
constraint, $(\pi^\mu_\rho)^2$, whereas the other components of the
constraint fix
the auxiliary fields in $X^-$.

The right-moving conformal anomaly can easily be shown to vanish
by adding the contributions from the four pairs of chiral
and anti-chiral bosonic superfields, $X^{+\bar l}$ and $X^{-l}$
(each pair contributes +3), from the two pairs of chiral
and anti-chiral fermionic superfields (although
$W^+$ and $W^-$ is not constrained to be chiral and
anti-chiral, the anti-chiral part of $W^+$ and the chiral part
of $W^-$ do not appear in the action, and
therefore do not contribute to the conformal anomaly),
$\Psi^\pm$ and $W^\pm$
(each pair contributes $-$3), and from the N=(2,0)
super-reparameterization ghosts (which contribute $-$6).

It is
interesting to note that in four and six spacetime dimensions,
all steps in the construction of the N=(2,0) action are
identical except that instead of four pairs of bosonic
superfields, there is only one pair in four dimensions and
two pairs in six dimensions. Because of this difference,
the conformal anomaly in these dimensions is equal to the conformal anomaly
of the Neveu-Schwarz-Ramond string in the same number of
dimensions (i.e., $-$9 in four dimensions and $-$6 in six dimensions).
In three spacetime dimensions, $\Dp\Theta^\alpha\gmu\Dp\Theta^\beta$
has no non-zero solutions, implying that there are no non-trivial
right-moving
solutions to the classical equations of motion.

The left-moving constraint is just $(\drhobar x^\mu)^2
+i\phi^p\drhobar\phi^p$ after using the equations of motion, and
generates the usual left-moving Virasoro algebra for the
heterotic superstring with no conformal anomaly (note that
the left-moving part of $x^+$ is treated differently from the
right-moving part).

With these N=(2,0) super-Virasoro constraints, it should be possible
to use standard BRST quantization techniques to calculate
scattering amplitudes$^{16}$ and to construct a second-quantized field
theory$^{36,37}$ for the heterotic Green-Schwarz superstring. Although the
BRST operator will not be manifestly covariant, it will still be
an improvement over the light-cone
method in which the BRST
operator is trivially zero since all of the gauge invariances
have been fixed.

Finally, it will be argued that the scattering amplitudes obtained
by integrating the exponential of the covariant action over all
N=(2,0) super-Riemann surfaces will lead to the same amplitudes as
those obtained using the new light-cone method (up to an as
yet undetermined measure factor).

After choosing N=(2,0) superconformal gauge, which allows the
action to be written in terms of free fields, the scattering
amplitude is:
$$\int dM_i dW^+dW^- d\Psi^+ d\Psi^- dX^{+\bar l} dX^{-l}
d\Phi^{+\bar q} d\Phi^{-q} ~(\lambda)^
{2g+n-2}~J~V_1(\rho_1)...V_n(\rho_n)$$
$$exp[
 \int d\tau d\sigma d\pp d\pd [{i\over 4}
(X^{+\bar l}\drhobar X^{-l}-X^{-l}\drhobar X^{+\bar l}) +
W^-\drhobar\Psi^+ -W^+\drhobar\Psi^- +{1\over 2}\Phi^{+\bar q}
\Phi^{-q}]$$
where $M_i$ are the super-moduli of the N=(2,0) super-Riemann surface
of genus g
with n punctures, $\lambda$ is the string coupling constant,
$J$ is the Jacobian that results from choosing N=(2,0) superconformal
gauge, $V_1...V_n$ are the n vertex operators,
and in addition to the chirality and
N=(2,0) super-Virasoro constraints, one has
global constraints coming from the requirement of equation (III.C.3c) that
$\Dp\Psi^+\Dm\Psi^--i\Psi^+\drho\Psi^-
-i\Psi^-\drho\Psi^+=\drho T$
for some real superfield $T$.

After integrating out the $W^\pm$ superfields
(assume that the vertex operators have been written in
light-cone gauge, so that they don't have any $W^\pm$ dependence),
one obtains the requirement that $\drhobar \Psi^\pm$=0.
This analyticity condition, when combined with the chirality constraints
and with the requirement of equation (III.C.3c),
restricts not only the range of integration for $\Psi^\pm$, but
also the range of integration for the N=(2,0) super-moduli, $M_i$.
In other words, only for a restricted class of N=(2,0) super-Riemann
surfaces does there exist single-valued fermionic superfields, $\Psi^\pm$,
which satisfy the above conditions (the other N=(2,0) super-Riemann
surfaces will not contribute to the scattering amplitudes). By
comparing with equation (II.B.2), one sees that this restricted
class contains precisely those N=(2,0) super-Riemann surfaces that can
be superconformally transformed into a light-cone interacting
super-worldsheet.

Since after integrating over $W^\pm$, the action is exactly
the light-cone action defined in equation (II.B.1), it only
remains to be shown that the Jacobian coming from the superconformal
gauge fixing cancels the Jacobian coming from expressing the
remaining N=(2,0) super-moduli
in terms of the light-cone parameters $\tilde\rho$, $\tilde\psi^\pm$,
and the twists. So up to this unknown measure factor, the scattering
amplitudes produced by the two different methods agree with each
other.

\vskip 24pt
\centerline {\bf Concluding Remarks}

It has been shown in this paper that by putting the heterotic
Green-Schwarz superstring on an N=(2,0) super-worldsheet,
rather than on an ordinary worldsheet, two major problems are
solved. In the light-cone method, integration over the super-moduli
allows the complicated light-cone interaction-point operators to
be removed, thereby eliminating the explicit dependence of the
scattering amplitude integrands on the locations of the interaction
points. In the covariant method, putting the action on an N=(2,0)
super-worldsheet allows an N=(2,0) superconformal gauge choice
in which the conformal anomaly vanishes and all fields are free.

It should also be much easier now to evaluate heterotic Green-Schwarz
superstring amplitudes, either using the new light-cone method on
N=(2,0) super-worldsheets or using BRST quantization in the
N=(2,0) superconformal gauge. It will be interesting to see how the
sum over spin structures in the Neveu-Schwarz-Ramond formalism is
replaced in the Green-Schwarz formalism (note that on an N=(2,0)
surface, the discrete spin structures of an N=1 surface are
replaced by the continuous U(1) moduli)$^{38}$. Perhaps the biggest
problem will be understanding how to describe the subset of
punctured N=(2,0) super-Riemann surfaces that correspond to
light-cone super-worldsheets.

Before closing, it should be remarked that the covariant heterotic
Green-Schwarz action on N=(2,0) super-worldsheets can be easily
generalized to describe coupling to background supergravity and
super-Yang-Mills fields, provided that the background fields
satisfy the usual on-shell supergravity and super-Yang-Mills
equations of motion.$^{18}$ This fact seems closely related
to the work of Howe, who has shown that the on-shell supergravity
and super-Yang-Mills equations of motion can be understood as
integrability conditions along ``pure-spinor'' strings in
loop superspace.$^{39}$

\vskip 24pt
\centerline {\bf Acknowledgements}
I would like to thank Jacques Distler, Michael Douglas, Jim Gates,
Murat Gunaydin, Paul Howe, Sergei Ketov, Stanley Mandelstam, Hiroshi
Nishino,
Ara Sedrakyan, Peter van Nieuwenhuizen, Ed Witten, and especially
Warren Siegel for useful discussions.
This work was supported by National Science Foundation grant
$\#$PHY89-08495.

\vfill\eject

\centerline{\bf References}

\item{(1)} Schwarz,J.H., Phys.Rep.89 (1982), p.223.

\item{(2)} Green,M.B. and Schwarz,J.H., Nucl.Phys.B243 (1984), p.475.

\item{(3)} Mandelstam,S., Prog.Theor.Phys.Suppl.86 (1986), p.163.

\item{(4)} Restuccia,A. and Taylor,J.G., Phys.Rev.D36 (1987), p.489.

\item{(5)} Mandelstam,S., Nucl.Phys.B64 (1973), p.205.

\item{(6)} Green,M.B. and Schwarz,J.H., Nucl.Phys.B243 (1984), p.285.

\item{(7)} Polyakov,A.M., Phys.Lett.B103 (1981), p.211.

\item{(8)} Carlip,S., Nucl.Phys.B284 (1987), p.365.

\item{(9)} Kallosh,R.E. and Morozov,A., Int.J.Mod.Phys.A3 (1988), p.1943.

\item{(10)} Kraemmer,U. and Rebhan,A., Phys.Lett.B236 (1990), p.255.

\item{(11)} Bastianelli,F., van Nieuwenhuizen,P., and Van Proeyen,A.,
Phys.Lett.B253 (1991), p.67.

\item{(12)} Mandelstam,S., Nucl.Phys.B69 (1974), p.77.

\item{(13)} Berkovits,N., Nucl.Phys.B304 (1988), p.537.

\item{(14)} Mandelstam,S., ``The n-loop Amplitude: Explicit Formulas,
Finiteness, and Absence of Ambiguities'', preprint UCB-PTH-91/53,
October 1991.

\item{(15)} Aoki,K., D'Hoker,E., and Phong,D.H., Nucl.Phys.B342 (1990),
p.149.

\item{(16)} Friedan,D., Martinec,E., and Shenker,S., Nucl.Phys.B271
(1986), p.93.

\item{(17)} Gross,D.J., Harvey,J.A., Martinec,E., and Rohm,R.,
Nucl.Phys.B256 (1985), p.253.

\item{(18)} Tonin,M., Phys.Lett.B266 (1991), p.312.

\item{(19)} Sorokin,D.P., Tkach,V.I., Volkov,D.V., and Zheltukhin,A.A.,
Phys.Lett.B216 (1989), p.302.

\item{(20)} Sin,S.J., Nucl.Phys.B313 (1989), p.165.

\item{(21)} Witten,E., ``D=10 Superstring Theory'', in {\it Fourth
Workshop on Grand Unification}, ed. P.Langacker et. al.,
Birkhauser (1983), p.395.

\item{(22)} Greensite,J. and Klinkhamer,F.R., Nucl.Phys.B291 (1987), p.557.

\item{(23)} Mandelstam,S., private communication.

\item{(24)} Ademollo,M., Brink,L., D'Adda,A., D'Auria,R.,
Napolitano,E., Sciuto,S., Del Giudice,E., DiVecchia,P., Ferrara,S.,
Gliozzi, F., Musto,R., Pettorini,R., and Schwarz,J., Nucl.Phys.B111
(1976), p.77.

\item{(25)} Cohn,J., Nucl.Phys.B284 (1987), p.349.

\item{(26)} Siegel,W., Phys.Lett.B128 (1983), p.397.

\item{(27)} Wiegmann,P.B., Nucl.Phys.B323 (1989), p.330.

\item{(28)} Sorokin,D.P., Tkach,V.I., and Volkov,D.V., Mod.Phys.Lett.A4
(1989), p.901.

\item{(29)} Berkovits,N., Nucl.Phys.B350 (1991), p.193.

\item{(30)} Penrose,R. and MacCallum,M.A.H., Phys.Rep.6C (1972), p.241.

\item{(31)} Berkovits,N., Nucl.Phys.B358 (1991), p.169.

\item{(32)} Brooks,R., Mohammad,F., and Gates,S.J.Jr., Class.Quant.Grav.3
(1986), p.745.

\item{(33)} Berkovits,N., Phys.Lett.B232 (1989), p.184.

\item{(34)} Henneaux,M. and Mezincescu,L., Phys.Lett.B152 (1985), p.340.

\item{(35)} Ivanov,E.A. and Kapustnikov,A.A., Phys.Lett.B267 (1991), p.175.

\item{(36)} Siegel,W., {\it Introduction to String Field Theory},
World Scientific (1988).

\item{(37)} Witten,E., Nucl.Phys.B268 (1986), p.253.

\item{(38)} Ooguri,H. and Vafa,C., Nucl.Phys.B361 (1991), p.469.

\item{(39)} Howe,P.S., Phys.Lett.B258 (1991), p.141.

\end